\documentclass[conference]{IEEEtran}
\IEEEoverridecommandlockouts
\usepackage{cite}
\usepackage{amsmath,amssymb,amsfonts}
\usepackage{algorithmic}
\usepackage{graphicx}
\usepackage{textcomp}
\usepackage{xcolor}
\usepackage{hyperref}
\def\BibTeX{{\rm B\kern-.05em{\sc i\kern-.025em b}\kern-.08em
    T\kern-.1667em\lower.7ex\hbox{E}\kern-.125emX}}
\begin{document}

\title{Case Study: Using AI-Assisted Code Generation In Mobile Teams}

\author{\IEEEauthorblockN{1\textsuperscript{st} Mircea-Serban Vasiliniuc}
\IEEEauthorblockA{\textit{Department of Computer Science} \\
\textit{Technical University Of Cluj-Napoca} and 
\textit{Nexttech International}\\
Cluj-Napoca, Romania \\
mircea.vasiliniuc@gmail.com}
\and
\IEEEauthorblockN{2\textsuperscript{nd} Adrian Groza}
\IEEEauthorblockA{\textit{Department of Computer Science} \\
\textit{Technical University of Cluj-Napoca}\\
Cluj-Napoca, Romania \\
adrian.groza@cs.utcluj.ro}
}

\maketitle

\begin{abstract}
The aim of this study is to evaluate the performance of AI-assisted programming in actual mobile development teams that are focused on native mobile languages like Kotlin and Swift.

The extensive case study involves 16 participants and 2 technical reviewers, from a software development department and it is designed to understand the impact of using Large Language Models trained for code generation in particular phases of the team, more specifically, technical onboarding and technical stack switch.
The study uses technical problems dedicated to each phase and requests solutions from the participants with and without using AI-Code generators. It measures time, correctness, and 'technical integration' using ReviewerScore, a metric specific to the paper and extracted from actual industry standards, the code reviewer of merge/pull requests.

The output is converted and analyzed together with feedback from the participants in an attempt to determine if using AI-assisted programming tools will have an impact on getting developers onboard in a project or helping them with a smooth transition between the two native development environments of mobile development, Android and iOS. The study was performed between May and June 2023 with members of the mobile department of a software development company based in Cluj-Napoca, with Romanian ownership and management.
\end{abstract}


\section{Introduction}
This paper has its motivation based on the actual requirements of a mobile development department from Cluj-Napoca, Romania which is specialized in native development. 
The need to understand how these LLMs can be tuned and used in real production projects and their potential impact on the productivity of such departments.

The characteristics of native mobile development involve programming in two main languages, Swift for iOS and Kotlin for Android. The study focuses on native development only and avoids implications with non-native technological stacks.

Two of the requirements of a mobile department lead are: (a) to create a technical onboarding procedure, and (b) to provide means for a developer to be proficient in the sibling platform (iOS to Android or vice-versa). 
The study is split following these two requirements. 
Having good procedures for these two phases ensures three major objective: (a) individual software engineering efficiency, (b) team collaboration regardless of platform, and (c) consistency in output between the two platforms. The latter is crucial for long-term projects.

In this line, the study aims to understand if using AI-assisted programming tools will have an impact on getting developers onboard in a project, helping them understand the sibling platform code better, or even helping them with a smooth transition between the two native development environments. The AI-assisted programming tools involved are Github Copilot and chatGPT. Mainly the study aims to answer the following research questions:
\begin{enumerate}
    \item[\emph{$RQ_1$}:] \emph{How can an AI-based code generator affect the experience when onboarding a new team member or switching technical stacks of an existing colleague?}
    \item[\emph{$RQ_2$:}] \emph{Can AI-based code generators affect the performance (completion time, correctness) of technical onboarding or technical stack switch tasks? }
    \item[\emph{$RQ_3$:}] \emph{Can AI-based code generators affect the technical integration efforts of a mobile development team?}
\end{enumerate}

The following sections will share studies performed in the past year related to the effort of understanding the impact of AI-Assisted code generation followed by the methodology of the study, the associated results, and the conclusions. 

\section{Related Work}
LLMs for Code, or Code Generation Models, are deep neural networks trained on large collections of code, mainly from open-source repositories. 
These models have achieved important results in several software engineering tasks like documentation generation, unit test generation, code generation, identification of issues, and even generating full functional programs based solely on natural language descriptions. AI assistants for coding tasks, powered by LLMs, are a relatively new set of tools. For example Codex~\cite{finnie2022robots}, the general-purpose programming model was released by Open AI in August 2021 and its public API was closed in March 2023 just to be fully integrated into GitHub Copilot~\cite{barke2023grounded, imai2022github}, which had its initial release in October 2021 and general release in June 2022. Other models, without open API or web interfaces like CodeBert~\cite{feng2020codebert} or AlphaCode~\cite{li2022competition},  were released in 2020 and 2022 respectively.

The studies related to the impact of these tools are mainly focused on Python-specific challenges.  
Vaithilingam et al.~\cite{b1} explored how programmers use and view Github Copilot, with 24 participants in a user study. 
The study's goal was to understand the different modes and perceptions of the tool's usage. 
The findings showed that even though Github Copilot did not improve the time or success rate of completing tasks, most participants preferred to have the tool as part of their regular programming activities.
This preference was due to Github Copilot often giving a useful initial point and reducing the need for online searches. However, participants faced difficulties in understanding, changing, and fixing the code snippets that Copilot produced, which greatly affected their efficiency in solving tasks. 

Kazemitabaar et al.~\cite{b2} focused on the usage of AI code generators to support novice learners in programming. 
The study argues that OpenAI Codex is an AI code generator that can help novice programmers by creating code from natural language descriptions. 
Still, it is also important to take into account the possible negative effects of depending too much on such tools for learning and remembering. 
A controlled experiment was done with 69 novices aged 10-17. The participants had to do 45 Python code-writing tasks, with half of them using Codex. After each code-writing task, a code-changing task was given. The results show that Codex helped a lot with code-writing performance, with a 1.15x higher completion rate and 1.8x better scores, without hurting performance on manual code-changing tasks. Moreover, learners who used Codex during the training phase did slightly better than their peers on evaluation post-tests done one week later, but this difference was not statistically significant.

Barke et al.~\cite{barke2023grounded} examined the ways users engage with an AI programming aide, particularly Github Copilot. The study involved observing 20 individuals as they utilized Copilot for various coding challenges. Some challenges involved adding to an existing codebase to better reflect genuine software development scenarios. These tasks encompassed several programming languages, such as Python, Rust, Haskell, and Java, to ensure no language preference. The methodology involved recording user interactions with Copilot, refining our observations into a cohesive theory, and tweaking the coding challenges based on emerging questions. 
Authors observed that some participants hoped to provide Copilot with a specific context, especially when detailing their work outside the study.

Our study aims to replicate industry-related procedures in order to extract the actual practicality of using such tools and to evaluate what can be improved to adopt or make use of AI-assisted code generation tools. 

\section{Designing the Study}
\begin{figure*}
\centerline{\includegraphics[width=0.77\linewidth]{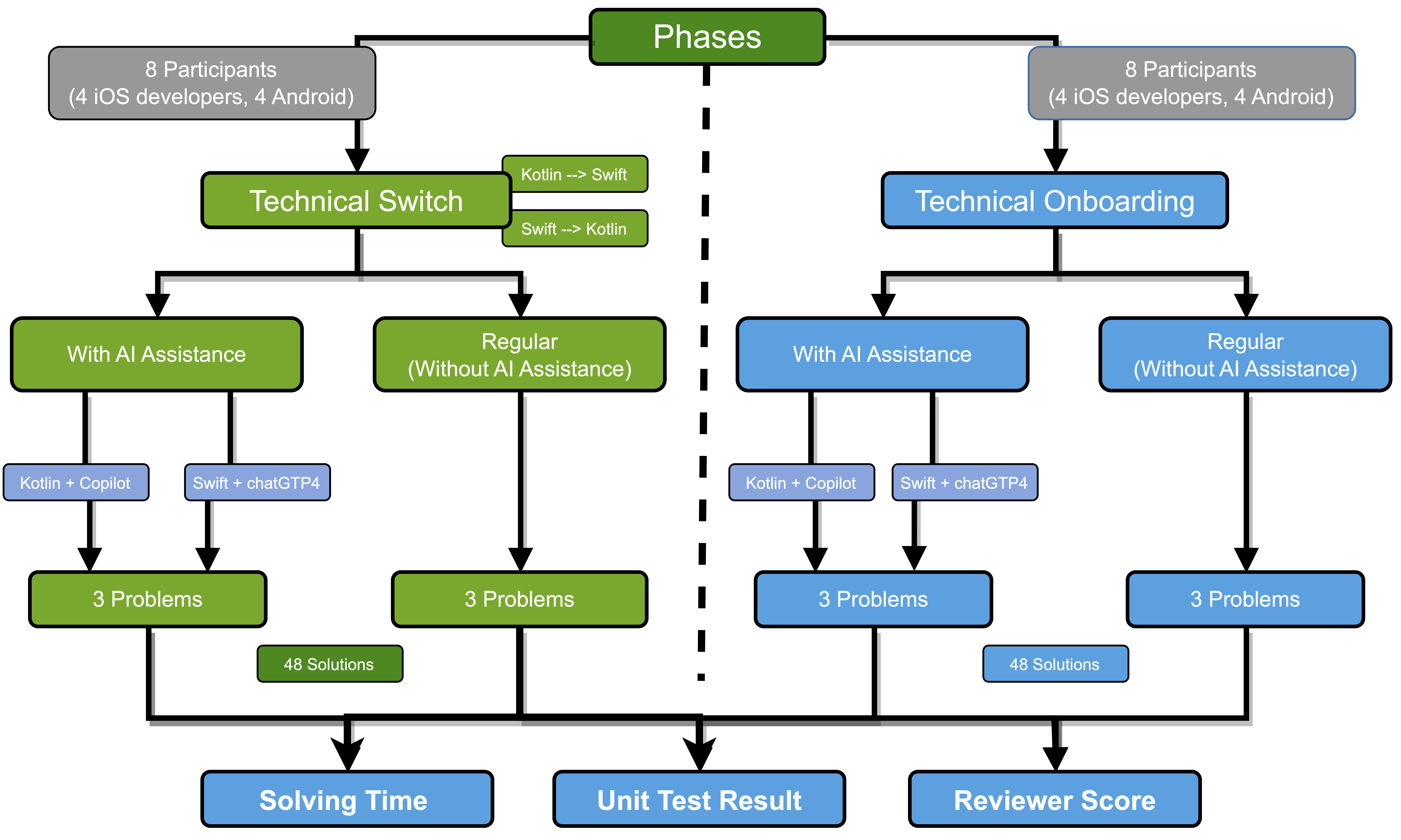}}
\caption{Breakdown of a participant output in the two phases 
}
\label{fig:caseStudy_Overview}
\end{figure*}

To analyze the impact of AI-assisted programming in a native mobile development environment, the study was designed to compare the output of development problems with and without the assistance of AI systems for each phase: technical onboarding, and technical stack switch. 
The study is separated into these two critical phases of a mobile team (Figure~\ref{fig:caseStudy_Overview}). Information about tasks, participants, measurements, and results will take into account each of the phases. 

When working on a native mobile project, in a sensitive sector (e.g. banking), a technical lead must prepare a team for projects with the following characteristics, the list not being comprehensive: 
 (1) The lifetime of the applications is long. Between 3 and 10 years;
 (2) The codebase must respect a quality gateway and technical and security audits;
 (3) The codebase must be verbose and highlight business-specific requirements, rules, and limitations; 
 (4) Knowledge transfer between team members must occur with minimum friction;
 (5) A high level of technical dependency on other services (environments, middlewares, backends, etc.);
(6) Constant changing or requirements due to legal or user-driven demands. Code must quickly respond to changes and extensions;
(7) Maintaining two codebases for Android and iOS applications.
These factors are pillars for the requirements of a mobile team and are used in the study design.
Note that the study can be considered a generic one, for any domain, with the observation that in sensitive projects, like banking, financial, or medical, there is a prioritization for having a standardized code base and a strong technical alignment within the team.

\subsection{Procedure}

\begin{figure}
\centerline{\includegraphics[width=\linewidth]{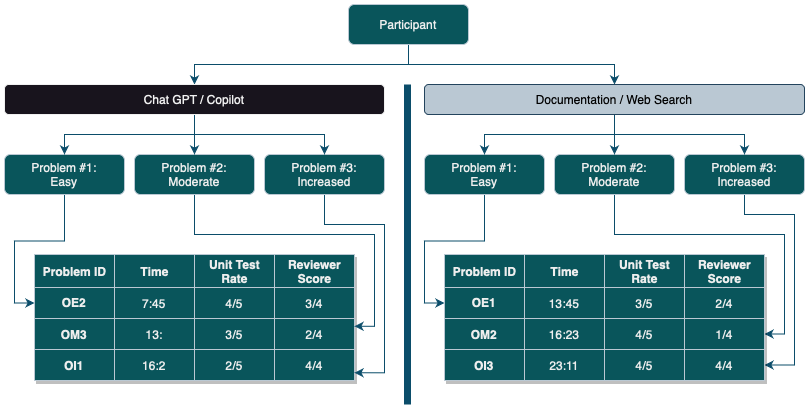}}
\caption{Breakdown of a participant output in any of the phases (Onboarding or Technical Switch)}
\label{fig:caseStudy_Participant}
\end{figure}

Each participant must solve two sets of problems. 
The first set contains one problem of each from the three difficulty levels (easy, moderate, increased) without using any AI-assisted tools with access to classic online resources (documentation, StackOverflow, search engines). This ensures the control conditions and the baseline required for the study (Figure~\ref{fig:caseStudy_Participant}).  
The second set also contains a problem of each difficulty but none of these problems are found in the first set. Participants must solve this second set using Github Copilot or chatGPT without access to anything else outside of documentation. 

The case study involved two main phases: Technical Onboarding and Technical Stack Switch. In the Technical Onboarding phase, the AI tool used for Kotlin was Github Copilot, while for Swift, it was chatGPT-4. There were a total of 9 distinct problems, 8 participants, and each participant worked on 6 problems. Out of the 6 problems, 3 were solved using the respective AI tool. In total, 48 solutions were generated, and each solution had 3 resulting metrics. The same AI tools were used in the Technical Stack Switch phase, and the metrics remained consistent (see Table~\ref{tab:caseStudy_overview_numbers}).

\begin{table}
\centering
\caption{Numbers involved in the Case Study}
\label{tab:caseStudy_overview_numbers}
\begin{tabular}{p{3.8cm}p{1.7cm}p{1.7cm}}
 & Onboarding & Stack Switch \\
\hline
AI Tool Used for Kotlin & Github Copilot & Github Copilot \\
AI Tool Used for Swift & chatGPT-4 & chatGPT-4 \\
No. of distinct problems & 9 & 9 \\
No. of participants & 8 & 8 \\
No. of problems per participant & 6 & 6 \\
No. of problems using AI Tool & 3 & 3 \\
No. of solutions generated & 48 & 48 \\
No. of metrics per solution & 3 & 3 \\
\hline
\end{tabular}
\end{table}

The investigation is conducted by utilizing an environment that possesses the following characteristics: 
First, tasks are resolved in a mobile-specific environment, Android Studio IDE for Android or Xcode for iOS. 
Second, for tasks that do not imply the usage of AI-assisted tools, the developer is free to use anything outside of these tools.
Third, for tasks that do imply the usage of AI-assisted tools, the developer 
can use only GitHub Copilot and ChatGPT. 
Fourth, each task is done under the supervision of a reviewer for gathering data, which includes the timing of each task and developer code writing behavior.
Fifth, the source files are collected by the reviewer 
Sixth, participants are requested to fill out a form for gathering qualitative results. 
Seventh, the time spent by each participant should not be more than 4 hours in total.

The metrics involved and the technical information associated that are integrated into the results are: 
\begin{itemize}
    \item \textit{Duration}. The time interval for solving the task. If time expires, the participant can no longer contribute and the solution is provided as it is.
    \item \textit{Correctness}. The unit tests are created by a reviewer based on problem-specific test cases. It aims to assess the correctness of the output and fitness of the function. For normalization factors, each problem has 5 test cases defined. These test cases aim for maximum correctness coverage. For each solution, actual unit tests are created from use cases and we retain the number of unit tests passed. Mainly, the score for each solution will be a figure from 0 to 5 representing the number of unit tests passed.
    \item \textit{Integration - ReviewerScore}. This consists of a qualitative marker that encapsulates the usage of platform APIs, the usage of most modern features of the language, and the verbosity of the code compared with the industry standards of the team which follows the Clean Coding principles (ref to Clean Coding book). Labels are Poor, Fair, Satisfactory, and Aligned (see Table~\ref{tab:ReviewerScore}).
\end{itemize}

\begin{table}
    \caption{The 4 levels or ReviewerScore}
    \label{tab:ReviewerScore}
    \centering
    \begin{tabular}{lp{6.5cm}}
        \textit{Poor} & the code violated all 3 features of the ReviewerScore.  \\ 
        \textit{Fair} & the code avoided at least one type of violation. This can be corrected using relatively low effort by a mentor \\
        \textit{Satisfactory} & the code violated only one type of criteria. \\ 
        \textit{Aligned} & No violations of the ReviewerScore features were found. 
    \end{tabular}
\end{table}

One extra type of information collected after the technical tasks are completed involves the filling of a form with a set of queries designed to extract user perception of the tasks. 
In this PostTaskSurvey (Table ~\ref{table:postTaskSurvey}), 
participants were asked to add their own observations for each item.

We predominantly focus on tasks that each span roughly 30 minutes. 
Our analysis, backed by statistical measures, displayed high confidence intervals for every metric. Such confidence levels underscore that our observed results are not merely incidental but represent meaningful patterns. A similar conclusion can be observed in Advait Sarkar et al.~\cite{b5}, where Co-Pilot was in focus. 
Moreover, while the core of our research revolves around these brief tasks, they bear significant implications for the broader industrial landscape. This is because when delving deep into industry challenges, many seemingly intricate issues can often be methodically broken down into tasks that range from 30 minutes to 4 hours. Because of this, our study doesn't just help us understand the small tasks we looked at but also gives clues about bigger issues in the industry. This can be really helpful for experts trying to solve tough problems in their fields. 
The analysis reveals a statistically significant association between the time taken to complete tasks, correctness of task completion, and technical alignment among the 16 participants ($p<.05$). 
This suggests that participants who use AI-Assisted tools tend to complete tasks faster but as correct and technically align as those without. However, the practical implications and specific nature of this relationship would require further investigation.

\begin{table}
    \centering
    \caption{The items of the Post Task Survey.}
    \begin{tabular}{p{0.5cm}|p{7cm}} 
    \textit{PTS1} & Rate the helpfulness of the AI-assisted tools (Github Copilot/Chat GTP) on a scale from 1 to 5. 1 means it didn't help in any manner, 5 meaning it solved the problem from the first interaction with it. \\  \hline
    \textit{PTS2} & Rate the level of understanding of the AI-assisted code on a scale from 1 to 5. 1 representing the participant didn't understand any instructions provided by the tool, and 5 meaning all instructions were clear. \\  \hline
    \textit{PTS3} & Rate the level of confidence you have in the final code on a scale from 1 to 5. 1 represents you aren’t confident, and 5 means full confidence that the code is functionally correct and respects the coding guidelines. 
    \\  \hline
    \textit{PTS4} & Rate the level of expected future usage of the AI-assisted code tools on a scale from 1 to 5. 1 being the participants don't see any future usage of the tools, 5 meaning that all future tasks will involve usages of AI-assistants. \\
    \end{tabular} 
    \label{table:postTaskSurvey}
\end{table}

\subsection{Tasks}

For a business area as demanding as banking and financing, mobile applications involve different levels of complexity and a variety of tasks. 
Developers are technically onboarded on a team, with a set of exercises that highlight these types of requirements. It is a means to communicate to them the expectations and to motivate them to be focused on constant improvements. The same aspects apply when assisting a developer to switch technical stacks 
between iOS and Android.

This study involves a simplification of what occurs in a standard technical onboarding process. In a real process, this procedure takes from 2 to 4 weeks, involves many more tasks, and implies providing solutions for small-scale projects or even full-featured implementations. The same applies to a technical stack switch, which usually involves gradual transition and procedures that can take from 1 to 3 months. 
The study attempts to simulate tasks given to developers involved in these situations, meant to help them achieve technical alignment with the team. In Table~\ref{table:studyTasksDifficulty}), we proposed three levels of difficulties, where the Time column indicates the time constraint for each type of level.

Since one of our goals is to specialize developers in the specifics of native mobile development, we require the usage of 'platform-specific representations' and APIs. This implies using the specialized type that the platform has provided for that representation. Examples: "abstract class Number" for Kotlin and "Numeric Protocol" in Swift for numbers, or "Date Structure" in Swift and "Date Class" in Kotlin.

\begin{table}
    \centering
    \caption{Difficulty levels for tasks involved in the study}
    \begin{tabular}{p{1.5cm}p{5cm}p{1cm}} 
        \textbf{Level} & \textbf{Description} & \textbf{Time (min.)} \\ \hline
        \textit{Easy} & Implying problems that do not assume expertise in the specific mobile platform. & 10. \\ \hline
        \textit{Moderate} & Implying problems that require some minimal research effort to be solved. & 20. \\ \hline
         \textit{Increased} & Implying problems that require a level of solution preparation or design. & 30. \\ \hline
    \end{tabular}
    \label{table:studyTasksDifficulty}
\end{table}

\subsubsection{Technical On-boarding}

In a software development department, mentorship starts with a proper technical onboarding procedure. This procedure usually involves setting up a set of tasks that help the new team member calibrate from a technical perspective, aligning him/her with expectations related to correctness and technical integration. Aspects like code structure, style, testability, and verbosity are verified and discussed in this process. 
Depending on the experience of the person who is being onboarded, this procedure can require a high effort of mentorship from an existing team member. In departments that are fully immersed in project-related tasks, it would be useful if some of the effort is 'delegated' to the AI assistants. This type of approach can also enhance the concept of a 'self-sufficient' developer which relies less on the team for tasks of medium to low complexity. 

This type of study and applicability is significant since the goal of any team is to achieve cost-effectiveness in developer onboarding.
AI-Assistants have the potential to guide the subject through the process with reduced feedback from mentors increasing independence and self-reliance and ideally knowledge.
One example of reduced feedback from a mentor is "Try to make the code more language idiomatic". 
This seems like an abstract request but with the help of an AI-Assistant that can modify or regenerate a piece of code to respect this, the results would ideally come faster.

One other challenge in the industry is the friction between approaches and communication between mentor and person onboarded. Not all developers react the same to a standard process of technical onboarding. The type of solutions presented in this paper can accommodate different ways of learning and provide some flexibility in the process. The aim is to reduce team efforts on the integration of a new member into the regular tasks of a team or department.

The last criterion is the scalability of the process. Can a Coding-Inteligence assistant help a team grow fast respecting a high bar related to technical onboarding? Therefore it is important for a mobile team to have efficient technical onboarding, for which AI-assisted programming can contribute.

We must specify that usually, technical onboarding tasks should focus on the usage of platform-specific APIs to encourage making the code resilient to SDK updates. Also, the technical onboarding tasks should revolve around the business sector that is the main focus of the department or team, specifically, the requirements should be similar in terminology and scope to the tasks usually occurring in a project. This would minimize the friction of the onboarded developer during the transition to an ongoing project.

\subsubsection{Technical Stack Switch}

It is estimated that as of 2020 there were 5,9 million Android developers and 2,8 million iOS developers in the world. If a mobile department is not capable of facilitating the transition between Android and iOS, it will lose ground compared to other departments. This doesn't necessarily apply only to the mobile sector. 
This concept can extend between the ReactNative technical stack and the native one (and vice versa) and also to web/frontend technical stack, like Angular, to native mobile stack. 
This research focuses only on the impact of AI assistance in the facilitation between the two major native mobile technical stacks.

There is a small number of mobile developers that can be proficient in both iOS and Android. 
The ones who do are of high importance to a mobile department focused on native applications in a complex domain like banking. This is due to their knowledge to facilitate a fundamental principle mentioned previously in the paper: having consistency between approaches in the two code bases that require maintenance. Therefore a goal of a mobile department is to have as many developers proficient in both platforms. Using AI-assisted programming in facilitating the switch can be highly productive.

Android and iOS SDKs, Kotlin and Swift, have strengths and weaknesses. By focusing on a good technical stack switch culture, a mobile team encourages a developer to be more versatile, and adaptable. 
In this type of switch, participants can adopt broader concepts in Software Engineering and increase their skills in the generic capability of such a profession: problem-solving. 
This phase assumes some knowledge of the mobile programming requirements. 
The aim is to reduce team efforts on the integration of members in a technical stack unknown to them before.

\subsection{Participants}

For each of the two areas of the study, we have 8 participants, having a total of 16 participants, assisted by 2 reviewers. Participants are members of the mobile department part of the Software Development company, situated in Cluj-Napoca, Romania. The department specializes in providing native applications and solutions, like frameworks or consultancy in sensitive areas like Banking and Financing. For both phases, 8 developers participated in the study, and for data collection, the procedure implied the usage of 2 reviewers.

\subsubsection{Technical On-boarding}
The participants are selected from the members that joined the mobile department in the last year. Only members with less than 3 years of experience participated, in order to avoid large impacts related to previous relative experience in the field. All participants finished a Computer Science studies, they  have basic knowledge in Software Development and they have previous interactions with mobile development SDKs. 
The two reviewers are Software Engineers who also share Technical Leadership responsibilities and they participate in the construction of technical onboarding procedures in the mobile department and provide consultancy in other software departments.
The participants were subject to the department's technical onboarding procedure when they joined the company but they didn't encounter the type of tasks used in the study.

\subsubsection{Technical Stack Switch}

The participants are selected from the active members in the mobile department. Only members with more than 3 years of experience on technical stack (Android or iOS) participated. 
All participants finished a Computer Science faculty, already accumulated industry experience, and are familiar with current industry and department standards. The same reviewers as in the Technical Onboarding were involved. All participants have minimal understanding of the sibling platform and never had official responsibilities to resolve tasks outside of their technology stack.

\section{Results}
First, we examine the metrics gathered in the two phases, by using the association between the 3 metrics: duration, correctness (unit test), and technical integration (ReviewerScore in Table~\ref{tab:ReviewerScore}) in association with the usage of AI tools, programming language, and problem complexity. 
Second, we draw conclusions related to the utility of the tools in a software development team specialized in mobile applications.

\subsection{Quantitative evaluation}

\begin{figure}
\centering
  \centerline{\includegraphics[width=0.9\linewidth]
  {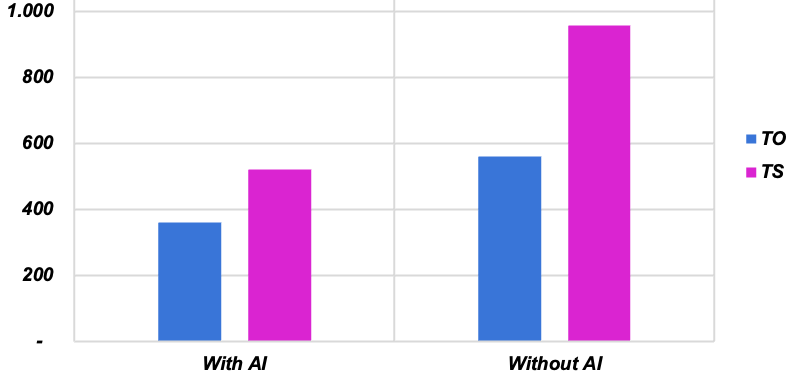}}
  \caption{Average time spent on tasks, regarding the the usage of AI tools. Time (y) is in seconds. TO = Technical Onboarding, TS = Technical Switch}
  \label{fig:caseStudy_time_avg}
\end{figure}

\paragraph{Time} Participants using AI Tools for Technical Onboarding Tasks had a 35\% decrease in the task duration and a 45\% decrease for Technical Switch tasks. 
We must reiterate that the study was designed to mimic challenges that are usually involved in mobile departments. For the Technical Switch phase, which contains problems that require knowledge in mobile development, the average time spent on a task was approximately 16 minutes, and with the usage of AI Tool, the average is lowered to 9 minutes (see Figure~\ref{fig:caseStudy_time_avg}). 

\begin{figure}
\centering
  \centerline{\includegraphics[width=\linewidth]
  {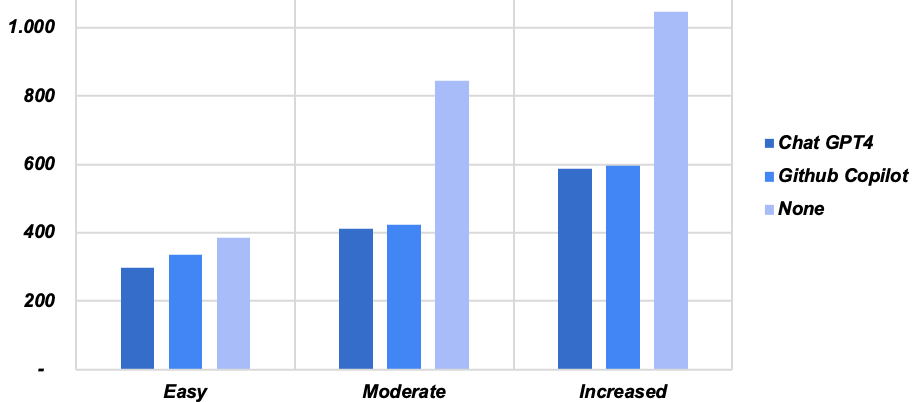}}
  \caption{Average time spent on tasks, for each difficulty level and with bars for each AI Tool, or None for solutions without AI-Tools. Time (y) is in seconds.}
  \label{fig:caseStudy_difficulty_time}
\end{figure}

For easy tasks, we observed little difference when AI-Tools were used and when not, but this changed for moderate where the time spent without an AI-Assistant was doubled on average, and for increased tasks, approx. 40\% more time was spent when participants didn't use AI-Tools and relied on regular sources like Google Search or StackOverflow besides the platform documentation. These differences can be seen in Figure~\ref{fig:caseStudy_difficulty_time}, which combines data from Onboarding and Technical Switch phases. 

\begin{figure}
\centering
  \centerline{\includegraphics[width=\linewidth]
  {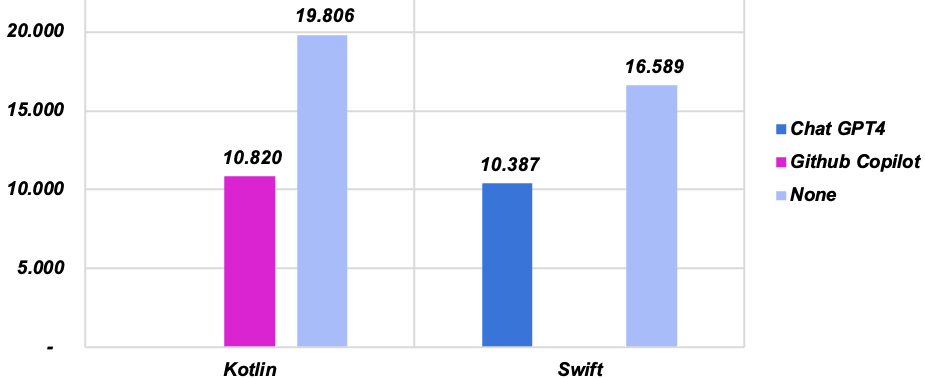}}
  \caption{Average time spent on tasks for each language and with bars for each AI Tool, or None for solutions without AI-Tools. Time (y) is in seconds.}
  \label{fig:caseStudy_language_time}
\end{figure}

There is a slight difference between the two platforms, iOS and Android, and their corresponding programming languages and AI Tools, ChatGPT4 and Github Copilot. The candidates onboarding or switching to iOS solved the tasks faster without assistance, one reason for this difference can be explained by the fact that, in the Technical Switch phase, the Android developers can adapt to iOS development faster than iOS can to the Android echo-system and the Kotlin language. No major difference was seen in the average time of the AI-Assistance usage with respect to the programming language and the corresponding AI-Tool, as seen in Figure~\ref{fig:caseStudy_language_time}.

\begin{figure}
\centering
  \centerline{\includegraphics[width=\linewidth]{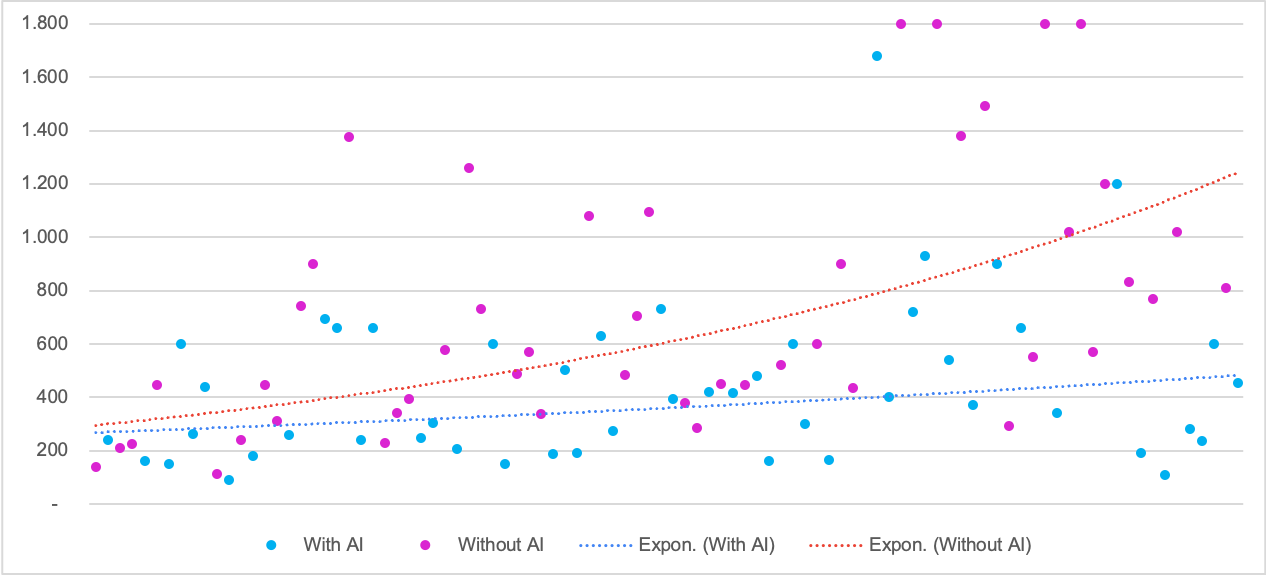}}
  \caption{Exponential trend-line in time for issues resolved with or without AI Tools. Time (y) is in seconds.}
  \label{fig:caseStudy_trendline_all}
\end{figure}

Finally, we noticed that without an AI Tool, the chances that a solution is not provided or provided after time expiration increased. 
As shown in Figure~\ref{fig:caseStudy_trendline_all} only in one case, a candidate crossed the 20-minute threshold using AI Tools and without tools, 8 solutions exceeded the threshold.

\begin{figure}
\centering
  \centerline{\includegraphics[width=0.9\linewidth]
  {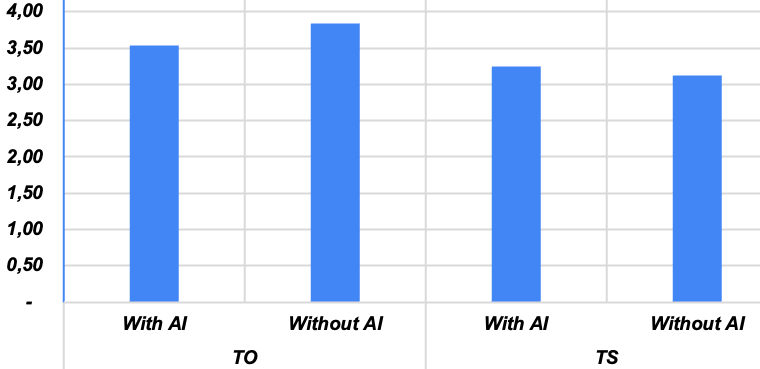}}
    \caption{Unit test score for all 4 categories of the study. Time (y) is in  seconds. TO = Technical Onboarding, TS = Technical Tech Stack Switch}
  \label{fig:caseStudy_unittest_avg}
\end{figure}

\paragraph{Correctness and Technical Integration} We measure correctness via unit tests and technical integration using the ReviewerScore in Table~\ref{tab:ReviewerScore}. 
On average unit tests for the task without AI-Tools had a higher score for the technical onboarding phase, 3.83/5, than with AI-Tools 3.54/5. 
We can assess that the majority of developers can evaluate the correctness risks and provide proper handling for it without the assistance of an AI tool (Figure \ref{fig:caseStudy_unittest_avg}), for tasks of low-medium complexity. 
A lower grade on AI-Tool can be explained by the decrease in ownership and understanding on the side of the candidate, in the solution generated using such a tool.

In the technical switch phase, unit test scores were higher for AI-Assisted solutions. 
This can be explained by the fact that models were already trained in the required programming language and can easier assess its particularities like nullability or error handling which can lead to lower correctness score. Developers who are new to a programming language might require more time to achieve this understanding.

\begin{figure}
\centering
  \centerline{\includegraphics[width=\linewidth]
  {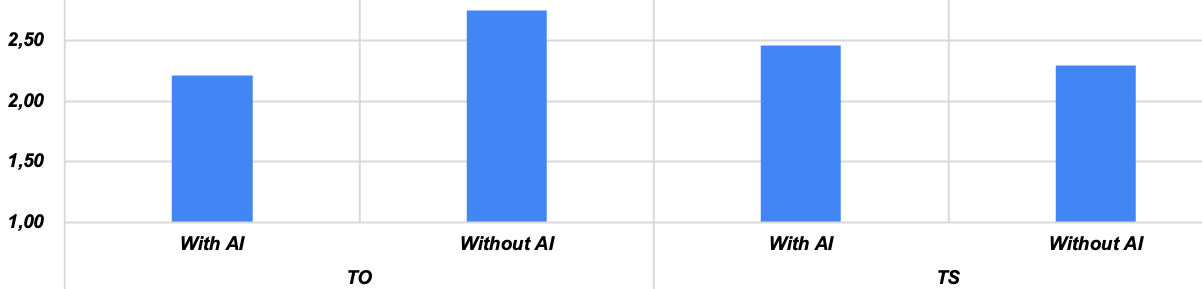}}
  \caption{ReviewerScore for all categories of the study. Time (y) is in seconds. TO = Technical Onboarding, TS = Technical Tech Stack Switch}
  \label{fig:caseStudy_reviewScore_avg}
\end{figure}

With respect to ReviewerScore, we have noticed a similar pattern as seen in Figure \ref{fig:caseStudy_reviewScore_avg}. On Technical Onboarding, a phase in which developers are already accustomed to the platform and programming language, solutions without AI-Assistance received better scores. On the other hand, the ones in Technical Switch favor AI-Tools usage, and this can be explained by the same reasoning, developers transitioning from one technology stack to another tend to be biased to the known stack and programming language and replicate approaches from that which are not in line with the standards of the new tech stack.

\begin{figure}
\centering
  \centerline{\includegraphics[width=\linewidth]
  {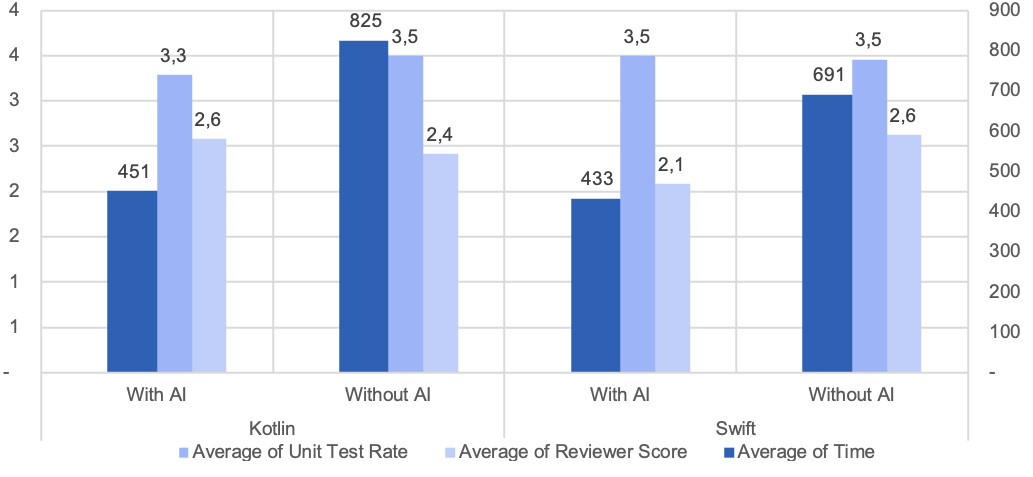}}
  \caption{Average Time, Average Unit Test Score, and Average Reviewer Score associated with mobile programming languages.}
  \label{fig:caseStudy_allMetrics_time}
\end{figure}

We gathered all qualitative metrics into one Figure \ref{fig:caseStudy_allMetrics_time} regardless of the type of problem, and based on the data supporting it. We can extract that time is significantly reduced in both mobile programming languages if AI-Assistance is used for code generation. Unit test scores have relatively minor differences, both Swift and Kotlin delta are approx. 1\%. ReviewerScore shows different variations based on the programming language, Swift showing a better score for solutions without AI Tools, and Kotlin solutions without AI-Tools having an 8\% increase.
 
\subsection{Qualitative evaluation}

\paragraph{Adopting the AI Tools} Candidates for technical stack switch had difficulties finding the correct sources to provide information about how to start, the APIs to be used, and language features due to a lack of experience in this type of data gathering. The sessions that implied AI had a quicker starting point since the role of these assistants is to aggregate data. 
Recall that this was a time-limited study. In a real technical switch or onboarding procedure, time is usually not a factor of pressure and developers do not need to quickly run after information, but tasks are usually more complex.

The candidates that had increased difficulties with the technical switch, represented by 0 unit test scores, or long times for solutions, had better results time-wise and correctness-wise with AI. 
Interactions with the ChatGPT4 were in general faster and provided more satisfaction than with the Github Copilot.

\paragraph{Quality Prompts} When working with a LLMs, providing appropriate prompts will increase the chances of a good result. An appropriate prompt will provide context and domain adaptation regarding the task at hand. 
In our case, we need to make sure the context is set to the corresponding programming language and the type of problem we need to solve. 

Another aspect is an expression of intent. For example, a quality prompt should contain elements related to the signature of a function and the expected result. Specificity is also a fundamental factor, which can be expressed by providing clear language types to be used in libraries or even design patterns. The problems in our case study were engineered with a low level of specificity, for two reasons, to avoid relying solely on AI tools and to mimic the industry procedure which explicitly avoids concrete details to evaluate the capacity of a developer to fill in the gaps in requirements with his/her own analytic capabilities. Candidates who were more attentive at modifying the prompts used in communication with the model have higher scores in correctness and ReviewerScore.

\paragraph{Incremental Prompts} A rewarding behavior extracted by the reviewers of the interactions with AI-Assisted tools was the incremental improvement of the solution by breaking the task into multiple prompts or altering the generated solution. Especially in the Technical Stack Switch, where candidates had already some experience on how to solve issues, solutions were improved by the "incremental prompts" strategy. 

\paragraph{Platform Association in Technical Stack Switch} We can assume the candidates of this phase know how to solve the received problems in the technical stack they have experience with. As reviewers, we observed if this knowledge benefits the overall result, especially when using AI-Tools. Candidates that had to solve Technical Switch tasks in Swift using the iOS SDK, had the 'capability' to interact with ChatGPT4 which can also provide natural language solutions related to code. A few candidates used this support to learn the equivalent of what they already knew on the platform in which they had experience. For the iOS / Swift developers who solved problems for this study using Android / Kotlin, the transition was more difficult because Copilot has reduced capability in association with languages outside of one of the projects. The experience of the candidates helped in this case since documentation can easily provide the API for a given need and afterward, this can be used in a more accurate prompt.

\begin{figure}
\centering
  \centerline{\includegraphics[width=\linewidth]
  {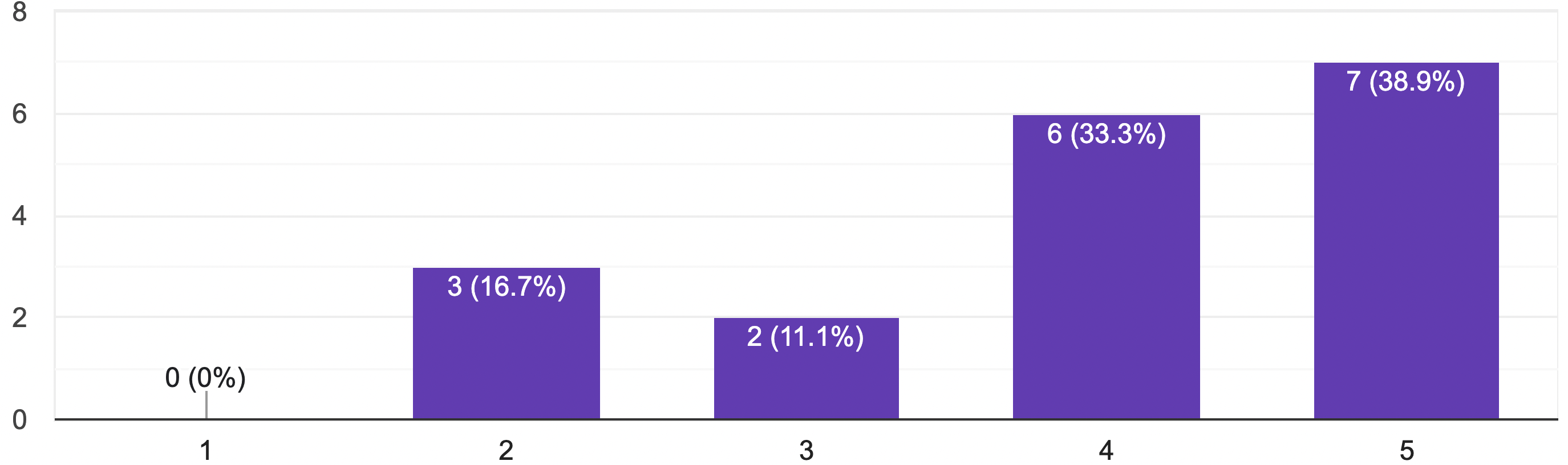}}
  \caption{Result to PTS1: Rate the helpfulness of the AI-assisted tools (Github Copilot/Chat GTP) on a scale from 1 to 5. 1 representing it didn't help in any manner, 5 meaning it solved the problem from the first interaction with it.}
  \label{fig:survey_pts01}
\end{figure}

\begin{figure}
\centering
  \centerline{\includegraphics[width=\linewidth]{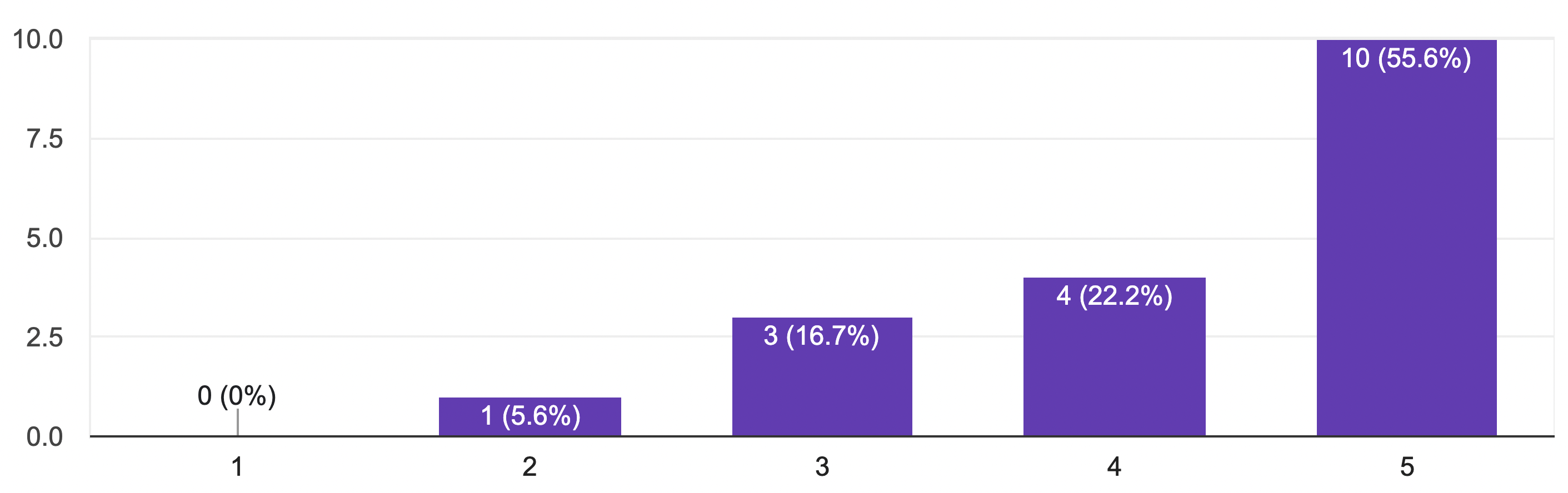}}
  \caption{Result to PTS2: Rate the level of understanding of the AI-assisted code on a scale from 1 to 5. 1 means the participant didn't understand any instructions provided by the tool, and 5 meaning all instructions were clear.}
  \label{fig:survey_pts02}
\end{figure}

\begin{figure}
\centering
  \centerline{\includegraphics[width=\linewidth]{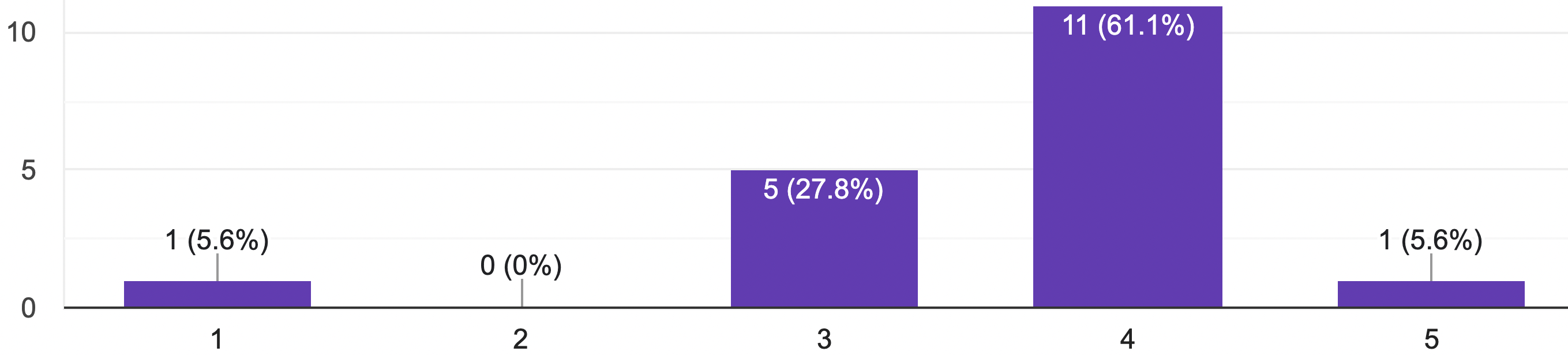}}
  \caption{Result to PTS3: Rate the level of confidence you have in the final code on a scale from 1 to 5. 1 means you aren’t confident, 5 means full confidence that the code is functionally correct and respects the coding guidelines} 
  \label{fig:survey_pts03}
\end{figure}

\begin{figure}
\centering
  \centerline{\includegraphics[width=\linewidth]{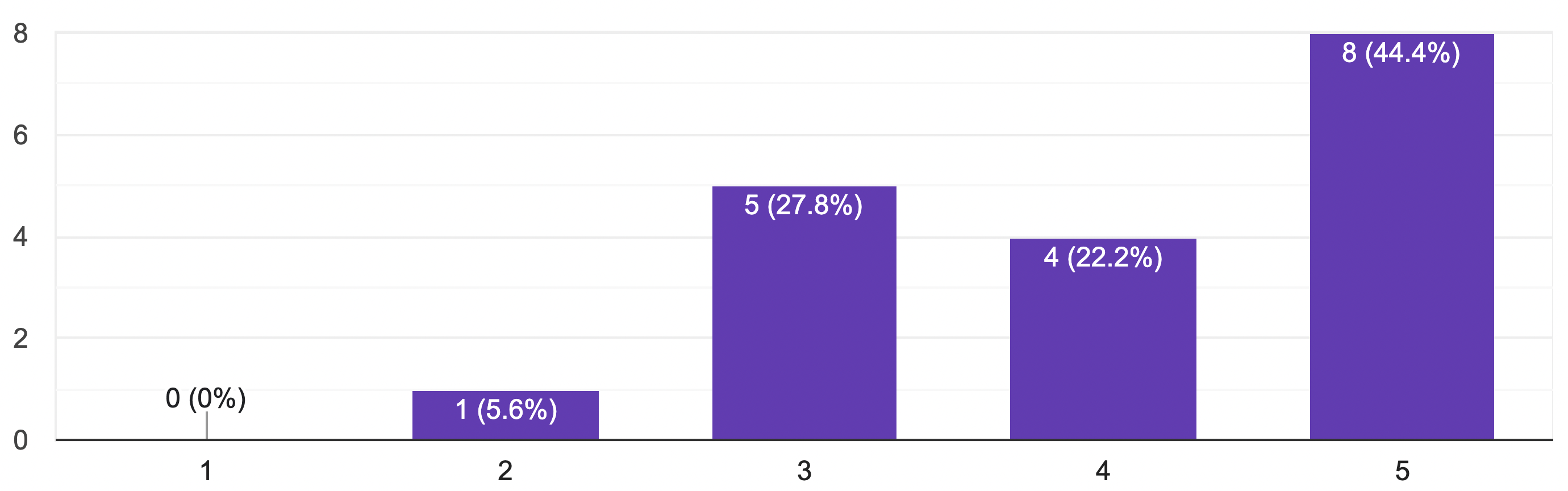}}
  \caption{Result to PTS4: Rate the level of confidence you have in the final code on a scale from 1 to 5. 1 meansyou aren’t confident, 5 means full confidence that the code is functionally correct and respects the coding guidelines} 
  \label{fig:survey_pts04}
\end{figure}

\paragraph{Post Study Participants Survey} From the survey that all the 18 participants were asked to fill out, 13 (72\%) of the participants rated the helpfulness of the AI-Tools as high (Figure~\ref{fig:survey_pts01}) and 14 of them (77\%) stated that they understood the solutions provided by the AI-Tools (Figure~\ref{fig:survey_pts02}). Regarding the confidence that the candidates had in the final output of the code, 12 (66\%) gave high confidence votes with 11 choosing second the maximum (Figure~\ref{fig:survey_pts03}). Of all the participants, 12 (66\%) stated that will rely on AI-Assisted code generation tools in the expected future, with the rest mostly being uncertain at this point (Figure~\ref{fig:survey_pts04}). Note also that Pearce et. al~\cite{b6} found that 40\% of the programs generated by Github Co-Pilot had (CWE) vulnerabilities.
 
\section{Conclusions}

\textit{How can an AI-based code generator affect the experience when onboarding a new team member or switching technical stacks of an existing colleague?} For these particular types of procedures in a mobile development team, meaning providing technical assistance for a new member or assisting a colleague to adopt the sibling technical stack, we can assess
based on the results of the study and the participant's feedback, that AI-Based code generators can improve the experience by providing an additional source of information which if combined with developer experience and analytic thinking can lead to rewarding results for both the subject of the procedure and the mentors assisting the procedure.

\textit{Can AI-based code generators affect the performance (completion time, correctness) of technical onboarding or technical stack switch tasks?} Results of the study reflect a clear improvement in the duration of achieving tasks, in a simulation of the two types of procedures, without major implications related to the correctness of the solutions provided. This can be seen as a positive effect on performance and it can be considered reasonable for mobile teams to adopt such solutions with the consideration that the long-term learning and information retention of the members must be measured before scaling the process to a standard one. The regular research, done before starting to write a solution, was reduced by the usage of AI tools, with some candidates doing API validation directly using the generated code, increasing the interaction with documentation.

\textit{Can AI-based code generators affect the technical integration efforts of a mobile development team?} The design of the study considers the notable factor of technical alignment with the requirements of a team, by the addition of the ReviewerScore, in order to measure the impact of integration efforts. 
Results show that the impact of using such tools can lower the alignment due to reliance on a tool and less on the team's context and requirement. 
Still, the risk can be taken into consideration when designing procedures for technical onboarding or technical stack switches. 
A key piece of information is that tools like Github Copilot can adapt to a specific project or team in time. Alternatives to increase chances of alignment are building a dedicated model, like IVA-CodeInt, a custom model built for generating platform-specific code for Swift and Kotlin, which can be tailored to the needs of the team or project. A team can also consider taking into solutions similar to the one presented by Shrivastava et. al~\cite{b3} in which the prompts take the context from the entire repository, thereby incorporating both the structure of the repository and the context from other relevant files (e.g. imports, parent classes).

\textit{Was prompt engineering relevant?}
Effective prompt engineering is an attribute that was again observed to affect the results and understanding of the problem and its solution in a significant manner.
The candidates who were more attentive at modifying the prompts used in communication with the model have higher scores in correctness and ReviewerScore. 
As observed by Zhou et. al~\cite{b4}, task performance depends significantly on the quality of the prompt, and most effective prompts have been handcrafted by humans.
The study reflected that transitioning from the iOS (Swift) technical stack to the Android (Kotlin) technical stack can require more effort with and without using the AI Code Generators. 
Candidates reported that GitHub Copilot has low support for detecting incorrectly generated code, and requires corrections from developers, while it is expected that the behavior of the tool increases with the understanding of the user and context.

\textbf{Authors Contributions:}
M. S. Vasiliniuc, Conceptualisation, Methodology, Software, Data Curation, Writing—original draft, and Writing—review and editing;
A. Groza, Conceptualisation, Methodology, Writing—review, and Supervision. 
The data that support this study are available \url{https://github.com/mvasiliniuc/iva-codeint-mobile}.

\textbf{Acknowledgment:} We thank the reviewers, the participants in the study and M. Joldos for configuring the DGX machine. 

\bibliographystyle{IEEEtran}
\bibliography{conference}

\end{document}